\documentclass[11pt]{article}
\usepackage{moriond,epsfig}

\def\Journal#1#2#3#4{{#1} {\bf #2}, #3 (#4)}

\def\NIM{\em Nucl. Instrum. Methods}

\def\PLB{{\em Phys. Lett.}  B}
\def\EPJ{{\em Eur. Phys.}  J.}

\def\PRD{{\em Phys. Rev.} D}

\def\JHEP{\em JHEP}

\def\be{\begin{equation}}
\def\ee{\end{equation}}
\def\bea{\begin{eqnarray}}
\def\eea{\end{eqnarray}}

\begin{document}
\vspace*{4cm}
\title{HEAVY FLAVOURS AT ZEUS AND H1}

\author{ S.K. BOUTLE }

\address{Department of Physics and Astronomy, University College London,\\
Gower Street, London WC1E 6BT, England}

\maketitle\abstracts{
An overview of recent measurements of charm and beauty production in ep collisions at HERA is presented. Various techniques are used by the ZEUS and H1 collaborations to efficiently tag heavy quarks in events and different regions of phase space are explored. Differential cross sections are measured in both photoproduction and deep-inelastic scattering. The predictions based on perturbative QCD calculations at next-to-leading order are generally found to describe the proton structure and the production of heavy quarks.}

\section{Introduction}
\label{intro}

Heavy quarks ($c$ and $b$ quarks) are produced at HERA prodominantly by the process known as boson-gluon fusion, where a photon emitted by the electron interacts with a gluon in the proton producing a $b\bar{b}$ or $c\bar{c}$ pair. This process is calculated using a convolution of the parton density function of the proton (non-perturbative), the parton scattering cross section (perturbative) and the fragmentation functions (non-perturbative), which describe the transition from the heavy quark to a meson.  Perturbative QCD calculations of the parton scattering process should be reliable since the virtuality $Q^2$ of the exchanged photon and the large mass of the produced quark, in the case of photoproduction, provide a hard scale. Hence, the study of heavy quark production at HERA is a stringent test of perturbative Quantum Chromodynamics (QCD). Cross section measurements for heavy quark production and the extraction of the charm and beauty contributions to the proton structure function $F_{2}$ will be presented and compared to theoretical QCD predictions.



The data used in these measurements were collected using the H1 \cite{h1det} and ZEUS \cite{zeusdet} detectors. These are multi-purpose detectors designed to study interactions in the HERA collider which, during the HERA I running period, collided protons at 820GeV with 27.5GeV electrons or positrons. After a luminosity upgrade in 2001, the proton energy was increased to 920GeV. During the upgrade period a silicon microvertex detector (MVD) was installed in the ZEUS detector. This detector component and the existing vertex detector in H1 enable precision heavy flavour measurements based on lifetime tagging to be made by both experiments. 

\section{Decay Tagging Methods}
\label{decay}

Charm and beauty decays can be tagged by identifying a lepton which is produced in the semileptonic decay of heavy quarks. Two variables can be used to discriminate between different quark decays. The first is the relative transverse momentum, $p_T^{\rm rel.}$, of the lepton with respect to the heavy flavour hadron which for experimental purposes is approximated to the direction of the associated jet. This variable can be used to discriminate between beauty and charm decays since the mass of the beauty quark is larger and therefore it has a harder $p_T^{\rm rel.}$ spectrum. The second variable is the signed impact parameter ($\delta$) which is defined as the transverse distance of closest approach of a track to the primary vertex. This variable reflects the lifetime of the quark and hence can be used to discriminate between charm and beauty decays and the decays of light quarks. The sign allows a statistical disentanglement of detector resolution effects from the effects of the decay lifetime of the heavy hadron. 

\section{Beauty and Charm in semileptonic decays}
\label{sec:muons}

\begin{figure}
\begin{center}
\psfig{figure=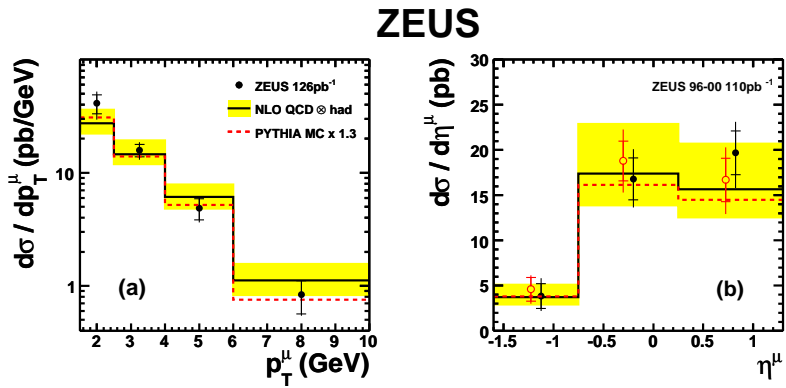,width=11cm}
\caption{Differential cross section as a function (a) $p_{T}^{\mu}$ and (b) $\eta^{\mu}$. The filled circles show the results from this analysis and the open circles show the results from the previous ZEUS measurement. The band represents the NLO QCD predictions with their uncertainties. The {\sc Pythia} MC predictions are also shown (dashed line).
\label{fig:bPHPmu}}
\end{center}
\end{figure}

The combination of the methods described in section \ref{decay} can be a powerful tool in the tagging of beauty quarks. Such a method was used by the ZEUS collaboration in a recent measurement \cite{ZEUS:beautyphp} of beauty photoproduction in the semileptonic decay channel into muons in dijet events (a similar measurement was carried out by the H1 collaboration \cite{H1:beautyphp}). Total and differential cross sections were measured in the kinematic region defined by $Q^2 < 1$ GeV$^2$, $0.2 < y < 0.8$, $p_T^{\rm jet 1,2} > 7,6$ GeV, $|\eta^{\rm jet}| < 1.5$, $p_T^\mu > 2.5$GeV ($p_T^\mu > 1.5$GeV for the $p_T^{\mu}$ cross section) and $-1.6 > \eta^\mu < 1.3$. Fig. \ref{fig:bPHPmu} shows the differential cross sections as functions of muon $p_T$ and $\eta$. The data are compared to a NLO QCD prediction computed with the FMNR program \cite{fmnr} and the $\eta$ distribution is also compared to a previous ZEUS measurement \cite{ZEUS:beautyold} which used the $p_T^{\rm rel.}$ method alone. Good agreement is observed with the calculations over the $p_T^\mu$ and $\eta^\mu$ regions considered and also with the previous measurement.

In a similar way, beauty and charm production has also been measured in DIS by the ZEUS collaboration \cite{ZEUS:beautydis}. In this analysis, it was also possible to extract beauty and charm contributions to the structure function, $F_{2}$.

\section{Inclusive track measurements}
\label{sec:incl}

An inclusive measurement of beauty dijets in the photoproduction regime carried out by the H1 collaboration \cite{H1:incltrack} is presented here. Photoproduction ($Q^{2}<1$~GeV$^{2}$, $0.15<y<0.8$) events with two jets with $p_{T}^{j1,j2}>11,8$~GeV and $-09.<\eta_{j1,j2}<1.3$ were selected. Events containing beauty quarks were distinguished from those containing only light quarks by reconstructing $\delta$ of the charged tracks, in a similar way to the muon analyses described in section \ref{sec:muons}.
The quantities $S_{1}$ and $S_{2}$, are defined as the significance $\delta/\sigma (\delta)$ of the track with the highest and second highest absolute significance respectively, where $\sigma(\delta)$ is the error on $\delta$. In order to eliminate a large fraction of the light quark background and to reduce the uncertainty due to the impact parameter resolution, the negative bins in the significance distributions were subtracted from the positive ones. To extract the beauty fraction, a simultaneous $\chi^{2}-$fit to the subtracted $S_{1}$ and $S_{2}$ distributions and to the significance of the reconstructed position of the secondary vertex ($\rm{L}_{xy}/\sigma(\rm{L}_{xy})$) was performed and used to calculate cross sections in $x$-$Q^2$ intervals. This could then be extrapolated to the full phase space to obtain the charm and beauty contributions to the structure function $F_{2}$.

\section{$D^{*}$ meson production}

\begin{figure}
\begin{center}
\psfig{figure=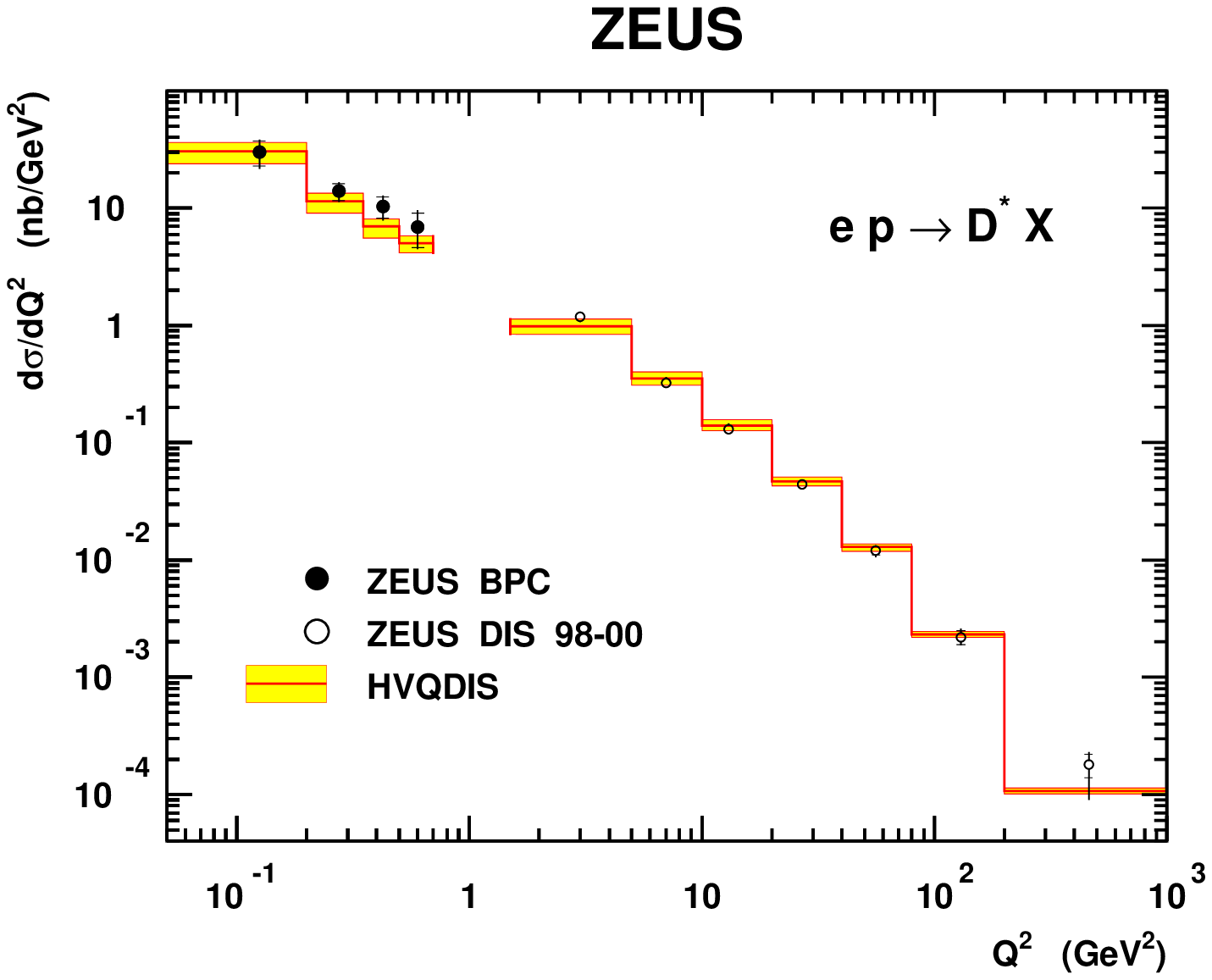,width=8cm}
\put(-200,170){\makebox(0,0)[tl]{\large (a)}}
\psfig{figure=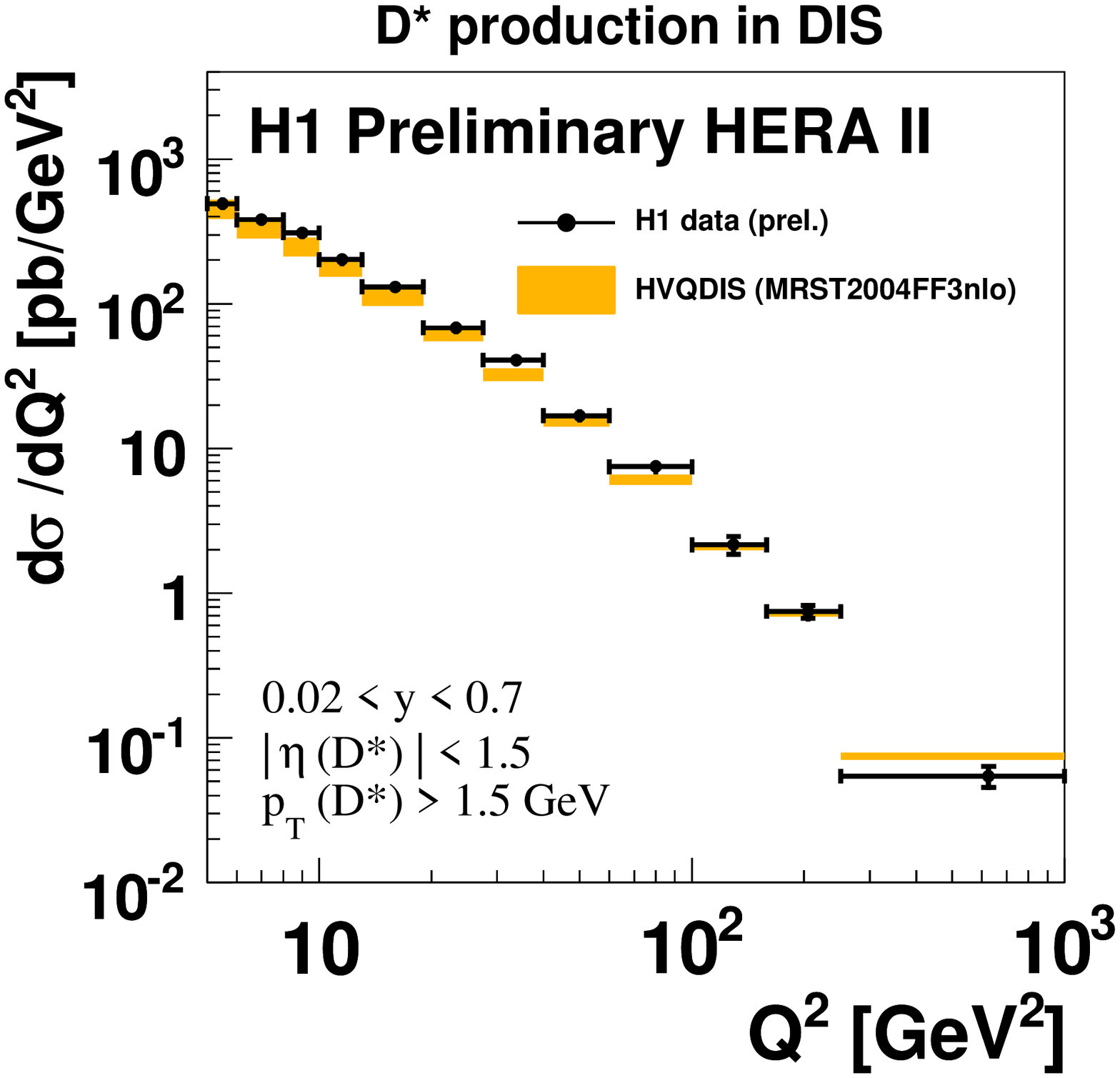,width=6cm}
\put(-180,170){\makebox(0,0)[tl]{\large (b)}}
\caption{Measurements of $D^{*}$ meson production cross section as a function of $Q^{2}$ performed by (a) the ZEUS and (b) H1 collaborations. The measurements are compared to theoretical predictions provided by the HVQDIS program.
\label{fig:dstar}}
\end{center}
\end{figure}

Charm production in DIS has been identified in the decay channel $D^{*\pm} \rightarrow D^{0}\pi^{\pm}_{slow} \rightarrow K^{\mp}\pi^{\pm}\pi^{\pm}_{slow}$. The $D^{*}$ cross section has been measured as a function of $Q^{2}$ by the ZEUS \cite{ZEUS:dstar} and H1 \cite{H1:dstar} collaborations as shown in Fig. \ref{fig:dstar}. The data are compared to NLO QCD predictions provided by the HVQDIS program \cite{hvqdis} and are well described by the calculation over four orders of magnitude. Using the same decay channel, $D^{*}$ meson production has also been studied in photoproduction by the H1 collaboration \cite{H1:dstarphp}. The theoretical uncertainties on these cross sections are much larger than the experimental uncertainties, indicating that higher order calculations are needed.

\section{Measurements of $F_{2}^{b\bar{b}}$ and $F_{2}^{c\bar{c}}$}

\begin{figure}
\begin{center}
\psfig{figure=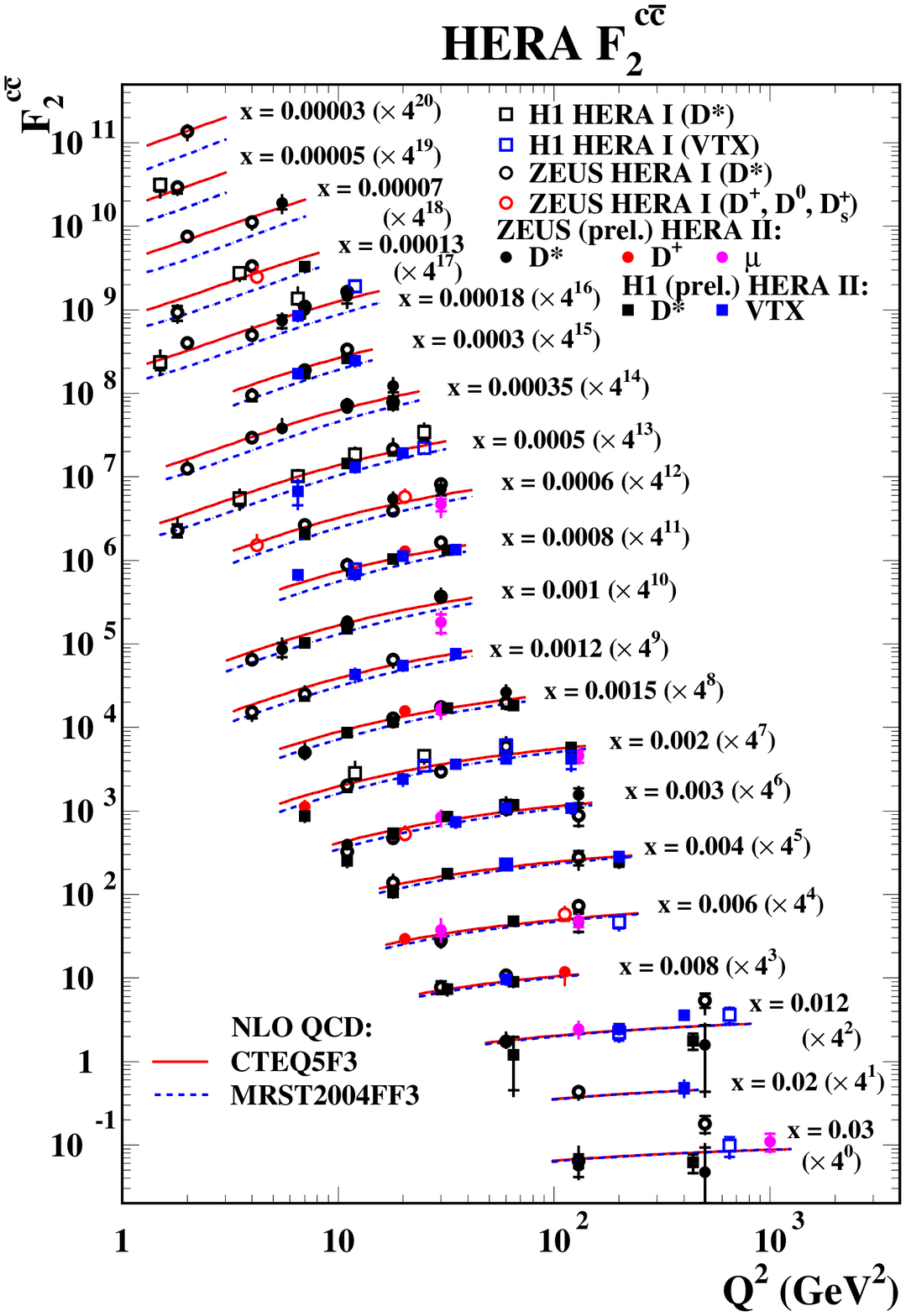,width=7cm}
\put(-200,230){\makebox(0,0)[tl]{\large (a)}}
\psfig{figure=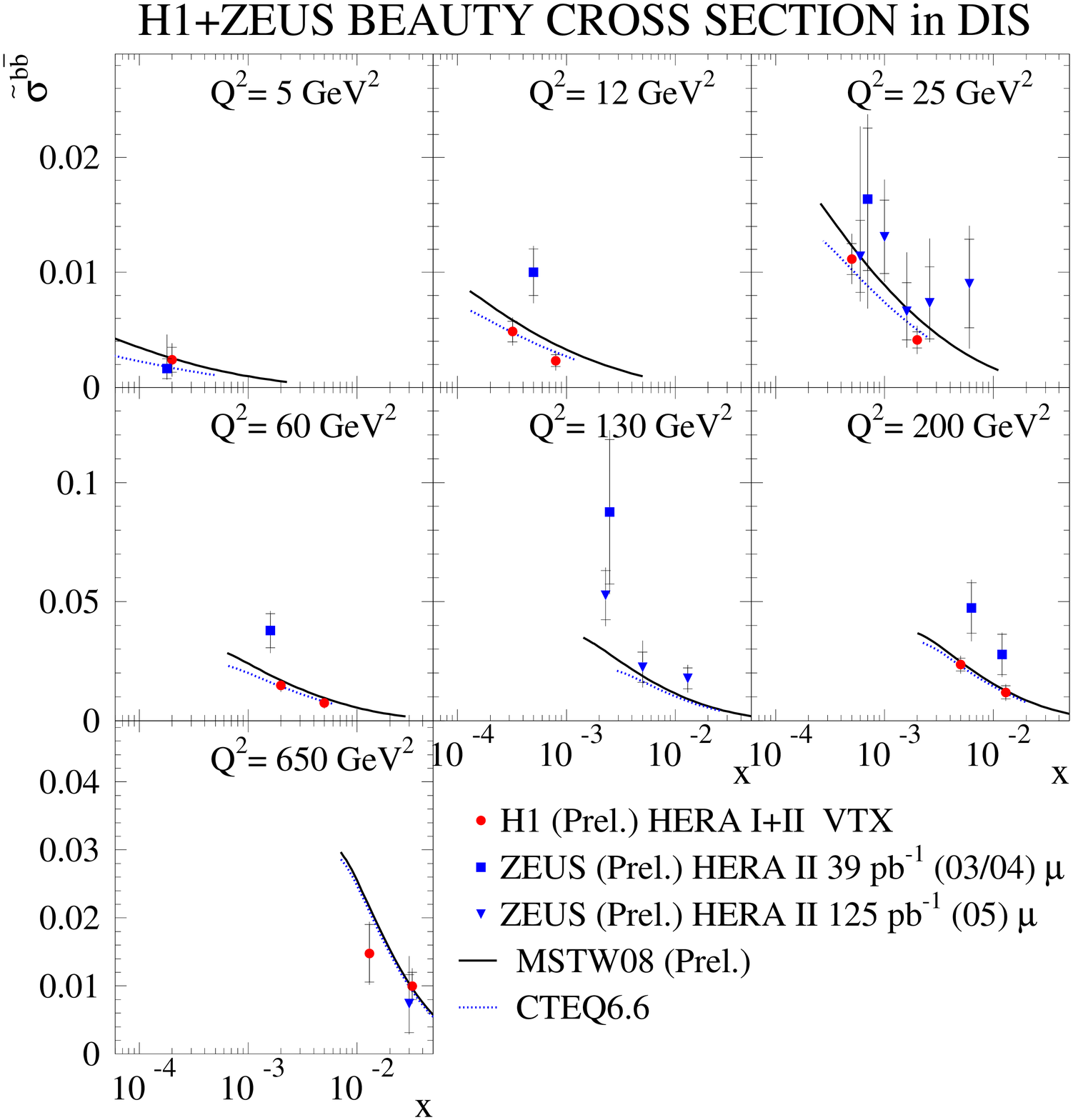,width=8cm}
\put(-220,230){\makebox(0,0)[tl]{\large (b)}}
\caption{(a) The structure function $F_{2}^{c\bar{c}}$ as a function of $Q^{2}$ for various values of $x$. (b) The reduced cross section $\tilde{\sigma}_{b\bar{b}}$ as a function of $x$ for different values of $Q^{2}$. The data are compared to QCD predictions.
\label{fig:f2cb}}
\end{center}
\end{figure}

Fig. \ref{fig:f2cb}(a) shows a summary of measurements of $F_2^{c\bar{c}}$ at fixed values of $x$ as a function of $Q^2$ measured using a variety of techniques while figure \ref{fig:f2cb}(b) shows measurements of $F_2^{b\bar{b}}$. The experimental points include those from the  analyses described in sections \ref{sec:muons} and \ref{sec:incl}, and in other H1 and ZEUS measurements. Good agreement is found between different data sets and analysis techniques.

The data are compared with QCD predictions from MRST \cite{mrst} and CTEQ \cite{cteq} at NLO. In general the results for $F_2^{c\bar{c}}$ are reasonably well described by the NLO predictions however at low $x$ there are notable differences between the two theoretical predictions. The precision on the data, in this case, is sufficiently high to be able to distinguish between different PDF sets. The same cannot be said for the measurements of $F_2^{b\bar{b}}$, which are limited by statistics.

\section{Conclusions}
\label{conc}

Recent results for charm and beauty production in $ep$ collisions at HERA have been presented and compared with NLO QCD calculations. By exploiting the precision of the vertex detectors of the H1 and ZEUS experiments new measurements have been made of beauty and charm cross sections and the charm and beauty contributions to the structure function $F_2$. These measurements have been well described by NLO QCD predictions. Measurements have been made by both collaborations of $D^{*\pm}$ production in the DIS regime which are described well by theoretical predictions.

\end{document}